\documentclass[amsmath,amssymb,twocolumn,prd,floatfix,showpacs, nofootinbib]{revtex4-1}
\usepackage{amsmath} 
\usepackage{graphicx} 
\usepackage{subfigure}
\usepackage{epstopdf} 
\usepackage{color} 
\usepackage{multirow}
\usepackage{setspace} 
\usepackage{overpic} 
\usepackage{amssymb}
\usepackage[normalem]{ulem}
\usepackage{natbib}
\usepackage[bookmarksnumbered, pdfstartview=FitH,colorlinks,urlcolor=blue, citecolor=blue,linkcolor=blue] {hyperref}
\usepackage{lineno}
\usepackage{bm}
\usepackage{rotating}
\usepackage{xcolor}
\usepackage{makecell}
\usepackage{mathtext}
\usepackage{mathrsfs}
\usepackage[utf8]{inputenc}
\usepackage[T1]{fontenc} 
\usepackage{tikz-feynman}  
\usetikzlibrary{arrows.meta}
\usetikzlibrary{decorations.pathmorphing}
\usepackage{adjustbox}
\let\oldequation\equation
\let\oldendequation\endequation
\renewenvironment{equation}
 {\linenomathNonumbers\oldequation}
 {\oldendequation\endlinenomath}

 \hypersetup{colorlinks = true,
            linkcolor = blue,
            urlcolor = blue,
            citecolor = blue,
            breaklinks=true,
            pdfstartview=Fit}

\newcommand{\BESIIIorcid}[1]{%
  \href{https://orcid.org/#1}{%
    \hspace{0.1em}%
    \raisebox{-0.45ex}{%
      \includegraphics[width=1em]{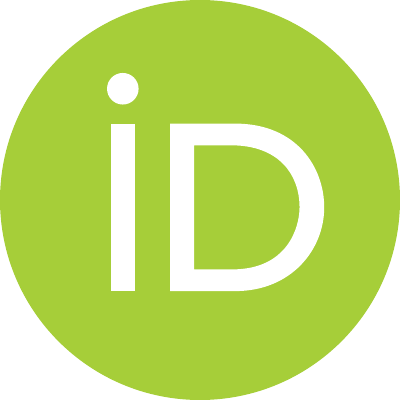}%
    }%
  }%
}

\begin{document}
\include{def-com-new}

\title{\boldmath Observation of the electromagnetic Dalitz decay $J/\psi \rightarrow \mu^+ \mu^- \eta$}

\author{
\begin{center}
M.~Ablikim$^{1}$\BESIIIorcid{0000-0002-3935-619X},
M.~N.~Achasov$^{4,c}$\BESIIIorcid{0000-0002-9400-8622},
P.~Adlarson$^{88}$\BESIIIorcid{0000-0001-6280-3851},
X.~C.~Ai$^{94}$\BESIIIorcid{0000-0003-3856-2415},
C.~S.~Akondi$^{33A,33B}$\BESIIIorcid{0000-0001-6303-5217},
R.~Aliberti$^{42}$\BESIIIorcid{0000-0003-3500-4012},
A.~Amoroso$^{87A,87C}$\BESIIIorcid{0000-0002-3095-8610},
L.~P.~An$^{54,h}$\BESIIIorcid{0000-0002-3274-5627},
Q.~An$^{83,70,\dagger}$,
M.~S.~Anderson$^{42}$\BESIIIorcid{0009-0008-1550-2632}
% ...
\\
\vspace{0.2cm}
(BESIII Collaboration)\\
\vspace{0.2cm} {\it
$^{1}$ Institute of High Energy Physics, Beijing 100049, People's Republic of China\\
% ...
$^{4}$ Budker Institute of Nuclear Physics SB RAS (BINP), Novosibirsk 630090, Russia\\
% ...
$^{33}$ INFN Sezione di Ferrara, (A)INFN Sezione di Ferrara, I-44122, Ferrara, Italy; (B)University of Ferrara, I-44122, Ferrara, Italy\\
% ...
$^{42}$ Johannes Gutenberg University of Mainz, Johann-Joachim-Becher-Weg 45, D-55099 Mainz, Germany\\
% ...
$^{54}$ Peking University, Beijing 100871, People's Republic of China\\
% ...
$^{70}$ State Key Laboratory of Particle Detection and Electronics, Beijing 100049, Hefei 230026, People's Republic of China\\
% ...
$^{83}$ University of Science and Technology of China, Hefei 230026, People's Republic of China\\
% ...
$^{87}$ University of Turin and INFN, (A)University of Turin, I-10125, Turin, Italy; (B)University of Eastern Piedmont, I-15121, Alessandria, Italy; (C)INFN, I-10125, Turin, Italy\\
$^{88}$ Uppsala University, Box 516, SE-75120 Uppsala, Sweden\\
% ...
$^{94}$ Zhengzhou University, Zhengzhou 450001, People's Republic of China\\

\vspace{0.2cm}
$^{\dagger}$ Deceased\\
% ...
$^{c}$ Also at the Novosibirsk State University, Novosibirsk, 630090, Russia\\
% ...
$^{h}$ Also at State Key Laboratory of Nuclear Physics and Technology, Peking University, Beijing 100871, People's Republic of China\\
% ...

}

\end{center}
\vspace{0.4cm}
}


\begin{abstract}
Using a data sample of $(8.998 \pm 0.039)\times10^9$ $J/\psi$ events collected by the BESIII detector at BEPCII, we report the observation of the electromagnetic Dalitz decay $J/\psi \to \mu^+ \mu^- \eta$ for the first time, with a statistical significance greater than 10$\sigma$. The branching fraction is measured to be $\mathcal{B}(J/\psi \to \mu^+ \mu^- \eta) = (3.05 \pm 0.14_{\rm stat.} \pm 0.13_{\rm syst.})\times 10^{-6}$.
The ratio of this branching fraction to that of the corresponding radiative decay $J/\psi \to \gamma\eta$ is determined to be $\mathcal{B}(J/\psi \to \mu^+\mu^-\eta)/\mathcal{B}(J/\psi \to \gamma\eta) = (2.80 \pm 0.18) \times 10^{-3}$.
Combining this result with the previous BESIII measurement of the $J/\psi \to e^+e^-\eta$ branching fraction yields $\mathcal{B}(J/\psi \to \mu^+\mu^-\eta)/\mathcal{B}(J/\psi \to e^+e^-\eta) = 0.219 \pm 0.020$.
These results are consistent with the theoretical predictions based on the vector meson dominance model.
\end{abstract}

\maketitle

\section{Introduction}
\label{int}
Electromagnetic (EM) Dalitz decays $V \to \ell^+\ell^- P$ provide a sensitive probe of the dynamic EM structure of the $V \to P$ transition 
and the underlying dynamics of photon–hadron interactions~\cite{Landsberg:1985fd}.
Here, $V$, $P$, and $\ell$ denote a vector meson, a pseudoscalar meson, and a charged lepton, respectively.
In the Standard Model (SM), the $\ell^+ \ell^-$ pair stems from an internal conversion of the virtual photon. 
If $V$ and $P$ are treated as point-like particles~\cite{Kroll:1955zu}, these EM Dalitz decays can be precisely described by quantum electrodynamics (QED) theory.
Otherwise, one needs to consider the modification caused by the dynamic EM structure arising at the vertex of the $V \to P $ transition, 
which is evaluated within quantum chromodynamics (QCD)~\cite{Achasov:1992ku,Klingl:1996by,Faessler:1999de,Terschluesen:2010ik,Ivashyn:2011hb}.

Normalized to the corresponding radiative decay $V \to \gamma P$, the partial decay width of $V \to \ell^{+} \ell^{-} P$ is given by 
\begin{equation}
\label{equ:dGVPll}
\begin{aligned} 
  \frac{\mathrm{d} \Gamma\left( V \rightarrow \ell^{+} \ell^{-} P \right)}{\mathrm{d} q^{2} \,\Gamma(V \rightarrow \gamma P )} 
  &= \left | F_{VP}(q^{2}) \right|^{2} \times {\rm QED}(q^{2}).
\end{aligned}
\end{equation}
Here, \(q\) is the four-momentum transfer, with \(q^2\) corresponding to the squared invariant mass of the \(\ell^+\ell^-\) pair (\(M^2_{\ell^+\ell^-}\)), 
and ${\rm QED}(q^{2})$ denotes the corresponding differential decay width calculated by QED~\cite{Kroll:1955zu} under the point-like particle hypothesis.
$F_{VP}(q^{2})$ is the normalized transition form factor (TFF) that characterizes the EM structure of the region in which $V$ converts into $P$, with the normalization of $F_{VP}(0)=1$. 
A major element governing $F_{VP}(q^{2})$ is the resonant interaction between photons and hadrons in the time-like region, commonly referred to as vector meson dominance (VMD). 
In the VMD model, the hadronic EM current is proportional to the vector meson fields~\cite{GellMann:1961tg, Bauer:1977iq}, 
and the TFF rises as a function of $M_{\ell^+ \ell^-}$.
With the single-pole approximation, the TFF can be parameterized as~\cite{ref:fu}
\begin{equation}
\label{equ:formfactor}
F_{V\!P}(q^{2}) = \frac{1}{1 - q^{2}/\Lambda^{2}},
\end{equation}
where $\Lambda$ is an effective pole mass that accounts for the overall effects of all possible resonance poles and scattering terms in the time-like kinematic region.

Recently, the VMD model underwent many experimental tests from EM Dalitz decays~\cite{pdg} measured by the A2, NA60, KLOE-2, BESIII, and LHCb Collaborations,
 such as $\pi^0 \to e^+ e^- \gamma$~\cite{pi0_to_gll_A2}, 
$\eta^{(\prime)}\to \ell^+ \ell^- \gamma$~\cite{eegamma_A2,mumugamma_NA60,mumugamma1_NA60,etaptoeegamma}, 
$\phi \to e^+ e^- \pi^0 /\eta$~\cite{pi0,eta1},
$J/\psi \to e^+ e^-  \eta/\eta(1405)/ X(1835)$~\cite{BAM:266,eta1405,X1835},
$h_c \to e^+ e^- \eta_c$~\cite{hc},
$\psi(3686)\to e^+ e^- \eta_c/ \chi_{cJ}$~\cite{etac,xcj},
$D^{*0} \to e^+ e^- D^0$~\cite{D0}, and 
$\chi_{c1} (3872)\to \mu^+ \mu^- J/\psi$~\cite{x3872_to_jpsimm_LHCb}. 
However, the TFF of the $\omega \to \mu^+\mu^-\pi^0$ decay~\cite{Arnaldi:2009aa,Arnaldi:2009aa1} measured by the NA60 Collaboration deviates from the VMD model by about a factor of 10 in the high $q^2$ region near the kinematic boundary.
This deviation cannot be adequately explained by other models~\cite{Terschluesen:2010ik, tershlusen_2012,schneider_2012}.
In addition, based on approximately 10 billion $J/\psi$ events,
the BESIII Collaboration observed a clear $\rho$-$\omega$ interference structure in the TFF of $J/\psi\to e^{+}e^{-}\pi^{0}$ as a function of $M_{e^{+}e^{-}}$~\cite{BAM:325}, 
which cannot be explained by the nonresonant VMD model~\cite{ref:fu,ref:21,ref:22}, effective Lagrangians~\cite{ref:23,ref:24}, or dispersion theories~\cite{ref:25}. After taking the $\rho$-$\omega$ interference structure into account, the measured branching fraction (BF) of $J/\psi\to e^{+}e^{-}\pi^{0}$~\cite{BAM:325} is twice the nonresonant VMD prediction~\cite{ref:22}.

These significant deviations motivate further studies of $V \to \ell^+ \ell^- P$ processes.
In this paper, we present the first experimental study of the EM Dalitz decay $J/\psi \to \mu^{+}\mu^{-}\eta$, 
by analyzing a $e^+ e^-$ collision data sample containing  $(8.998\pm 0.039)\times10^{9}$ $J/\psi$ events~\cite{Ablikim:Jpsi} collected at the center-of-mass energy $\sqrt{s}=3.097~\mathrm{GeV}$ in 2009, 2017, and 2018 with the BESIII detector.
The BF predicted by the VMD model~\cite{ref:fu} is $\mathcal{B}(J/\psi \to \mu^+ \mu^- \eta) = (0.30\pm0.01)\times 10^{-5}$, where the TFF is parameterized by Eq.~(\ref{equ:formfactor}) with $\Lambda =3.686$ GeV/$c^2$. Comparing it with the measured BF provides a test of the VMD model. In addition, combining our measurement with the existing measurement of $J/\psi \to e^+e^-\eta$~\cite{BAM:266} allows us to determine the ratio of the BFs of the two decay modes, providing a test of lepton flavor universality. Furthermore, the muonic channel provides complementary information on the $J/\psi\to\eta$ transition dynamics in a different $q^{2}$ region.

\section{BESIII DETECTOR and MONTE CARLO SIMULATION}
\label{sec:BES}
The BESIII detector~\cite{24} records symmetric $e^{+}e^{-}$ collisions provided by the BEPCII storage ring~\cite{25} in the center-of-mass energy range from 1.84 to 4.95~GeV, with a peak luminosity of $1.1 \times 10^{33}~\text{cm}^{-2}\text{s}^{-1}$ achieved at $\sqrt{s} = 3.773$~GeV.
BESIII has collected large data samples in this energy region~\cite{26,27,28}. The cylindrical core of the BESIII detector covers 93\% of the full solid angle and consists of a helium-based multilayer drift chamber (MDC),
a plastic scintillator time-of-flight system (TOF), and a CsI(Tl) electromagnetic calorimeter (EMC), which are all enclosed in a superconducting solenoidal magnet providing a 1.0~T magnetic field.
The solenoid is surrounded by the muon counter (MUC), which is an octagonal flux-return yoke made of steel, interleaved with resistive-plate-counter muon-identification modules. 
The charged-particle momentum resolution at $1~{\rm GeV}/c$ is $0.5\%$, and the resolution of the specific energy loss ${\rm d}E/{\rm d}x$ in the MDC is $6\%$ for electrons from Bhabha scattering. 
The EMC measures photon energies with a
resolution of $2.5\%$ ($5\%$) at $1$~GeV in the barrel (end cap) region. The time resolution in the TOF barrel region is 68~ps, while
that in the end cap region was 110~ps. The end cap TOF system was upgraded in 2015 using multigap resistive plate chamber technology, providing a time resolution of 60~ps, which benefits 98\% of the data used in this analysis~\cite{etof1}. 
The MUC is divided into eight and four segments in the barrel and end cap regions, respectively. The total thickness of the absorber material (iron) is approximately 45 cm, including the coil and the yoke. The typical rate at which pions are misidentified as muons is about 1-2\%, while the detection efficiency for muons with momenta above 0.4~GeV/$c$ is better than 95\%.

Monte Carlo (MC) simulated data samples produced with a {\sc
geant4}-based~\cite{geant4} software package, which
includes the geometric description of the BESIII detector and the detector response, are used to determine detection efficiencies and to estimate backgrounds. The simulation models the beam energy spread and initial state radiation in the $e^+e^-$ annihilations with the generator {\sc
kkmc}~\cite{ref:kkmc}.
The inclusive MC sample includes both the production of the
$J/\psi$ resonance and the continuum processes incorporated in {\sc
kkmc}~\cite{ref:kkmc}.
All particle decays are modeled with {\sc
evtgen}~\cite{ref:evtgen} using BFs either taken from the
Particle Data Group (PDG)~\cite{pdg}, when available,
or otherwise estimated with {\sc lundcharm}~\cite{ref:lundcharm}.
Final state radiation from charged final state particles is incorporated using the {\sc
photos} package~\cite{photos2}.%
 
Taking into account the polarization of the $J/\psi$ produced in $e^+ e^- $ collisions~\cite{ref:24}, we generate the signal $J/\psi\to\mu^{+}\mu^{-}\eta$ decay according to the differential decay width described by Eq.~(\ref{equ:dGVPll}) with the TFF parameterized as in Eq.~(\ref{equ:formfactor}). 
Here the pole mass $\Lambda = (2.56\pm 0.04_{\rm stat.} \pm 0.03_{\rm syst.})$ GeV/$c^2$ in the TFF is taken from the previous BESIII measurement of the $J/\psi \to e^+ e^- \eta$ decay~\cite{BAM:266},
since the EM structure at the vertex of $J/\psi \to \eta$ does not depend on the lepton flavor in the SM. 
The $\eta \to \gamma\gamma$ decay is subsequently generated uniformly in phase space.

\section{EVENT SELECTION criteria}
\label{sec:selection}
The decay chain of interest is $J/\psi\to\mu^{+}\mu^{-}\eta$, $\eta \to \gamma \gamma$. 
Charged tracks detected in the MDC are required to be within a polar angle ($\theta$) range of $|\rm{cos\theta}|<0.93$, 
where $\theta$ is defined with respect to the symmetry axis of the MDC, 
and to have a distance of closest approach to the interaction point (IP) of less than $\pm10$~cm along the beam direction and less than 1~cm in the plane transverse to the beam direction. The number of good charged tracks per event is required to be two with a net charge of zero.

Photon candidates are identified using isolated showers in the EMC. 
The deposited energy of each shower must be more than 25~MeV in the barrel region ($|\cos \theta|< 0.80$) and more than 50~MeV in the end cap regions ($0.86 <|\cos \theta|< 0.92$).
To exclude showers that originate from
charged tracks, the angle subtended by the EMC shower and the position of the closest charged track at the EMC must be greater than 10~degrees as measured from the IP.
To suppress electronic noise and showers unrelated to the event, the difference between the EMC time and the event start time is required to be within 
[0, 700]\,ns. At least two good photons are required.

\begin{figure}[t]
    \parindent=0pt
    \centering
    \includegraphics[width=0.46\textwidth]{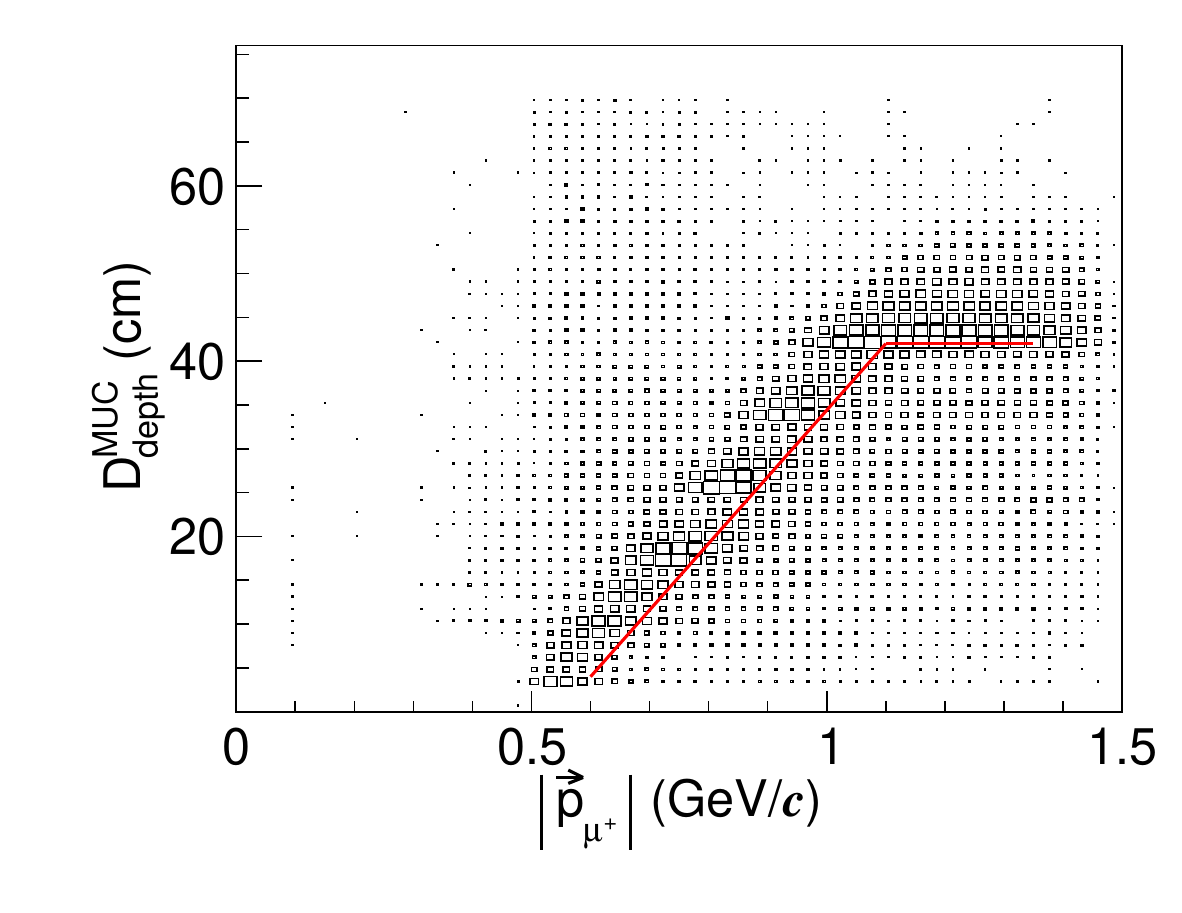}
    \setlength{\abovecaptionskip}{-1pt}
    \caption{Momentum-dependent penetration depth for $\mu^+$ candidates in the signal $J/\psi \to \mu^{+} \mu^{-} \eta$ MC sample. The red line denotes the requirements listed in Table~\ref{tab:depth}.}
    \label{fig:depth}
\end{figure}

Particle identification (PID) for charged tracks combines measurements of the specific ionization energy loss in the MDC~(d$E$/d$x$), 
the flight time in the TOF, and the energy deposited in the EMC ($E_{\rm EMC}$), to form likelihoods ($\mathcal{L}$) under the muon and positron (electron) hypotheses. 
Tracks are identified as muons when they satisfy $\mathcal{L}(\mu) > \mathcal{L}(e)$.
To further suppress pion misidentification, each muon candidate is required to satisfy $0.16 \leq E_{\rm EMC} \leq 0.22$ GeV and to have hits in at least five MUC layers.
Table~\ref{tab:depth} summarizes the requirements on the MUC penetration depth ($D_{\rm depth}^{\rm MUC}$) as a function of the muon momentum ($|\vec{p}_{\mu^{\pm}}|$), 
which are illustrated in Fig.~\ref{fig:depth}.
MC studies show that, after applying these selection criteria, 53.5\% of the signal events are retained, while about 98.2\% of the $J/\psi \to \pi^+\pi^-\eta$ background events are rejected.

\begin{table}[hbpt]
    \centering
    \caption{Momentum-dependent requirements on the penetration depth in the MUC.}
    \renewcommand{\arraystretch}{1.2}
    \setlength{\extrarowheight}{1.2ex}
    \setlength{\tabcolsep}{3.5pt}
    \renewcommand{\arraystretch}{1.0}
    \vspace{0.2cm}
        \begin{tabular}{ccl }
        \hline\hline
        $\lvert \vec{p}_{\mu^{\pm}}\lvert$ (GeV/$c$) & 
        $D_{\text{depth}}^{\text{MUC}}$ (cm) \\
        \hline
        (0.0, 0.6)   & No requirement \\

        [0.6, 1.1)   &   $ > (76\cdot \lvert \vec{p}_{\mu^{\pm}}\lvert -41.6)$  \\

        [1.1, 1.5)   &$ > 42$   \\
        \hline\hline
        \end{tabular}%
    \vspace{-0.1cm}
    \label{tab:depth}
\end{table}

A four-constraint (4C) kinematic fit imposing overall energy-momentum conservation is performed on the $\mu^{+}\mu^{-}\gamma\gamma$ combinations, 
and  events with $\chi^{2}_{\rm 4C}<20$ are retained. 
This requirement is optimized by maximizing the figure of merit (FOM), defined as
FOM = $S/\sqrt{S+B}$, 
where \(S\) and \(B\) are the numbers of expected signal and background events estimated from the signal and inclusive MC samples, respectively. 
For events with more than two photon candidates, the combination with the smallest $\chi^{2}_{\rm 4C}$ is kept.
The $\eta$ candidate is reconstructed from a $\gamma \gamma$ pair with an invariant mass $M_{\gamma \gamma}$ in the range $[0.52, 0.58]~{\rm GeV}/c^2$. 
Further selection criteria are based on the four-momenta updated by the kinematic fit.

Due to pion misidentification, 
the $J/\psi \to \pi^+\pi^-\eta$ background can mimic the signal.
MC studies show that about 67\% of $J/\psi \to \pi^+ \pi^- \eta$ background events are rejected by the requirement $\chi^2_{\rm 4C}(\mu^+ \mu^-\eta)<\chi^2_{\rm 4C}(\pi^+ \pi^- \eta)$,
where $\chi^2_{\rm 4C}(\pi^+ \pi^- \eta)$ is obtained under the assumption that both good charged tracks are pions.
Additionally, the variable $(|E_{\gamma_1} - E_{\gamma_2}|)/|\vec{p}_{\eta}|$ is required to be less than 0.8 to reject 
the $J/\psi \to \gamma\eta^{\prime} (\to \pi^+ \pi^- \gamma)$ background, where $E$ and $|\vec{p}|$ are the energy and momentum of the corresponding particle.
 
\begin{figure}[htbp]
   \parindent=0pt
   \centering
   \includegraphics[width=0.48\textwidth,
    trim={0cm 2.50cm 0cm 0cm},
    clip ]{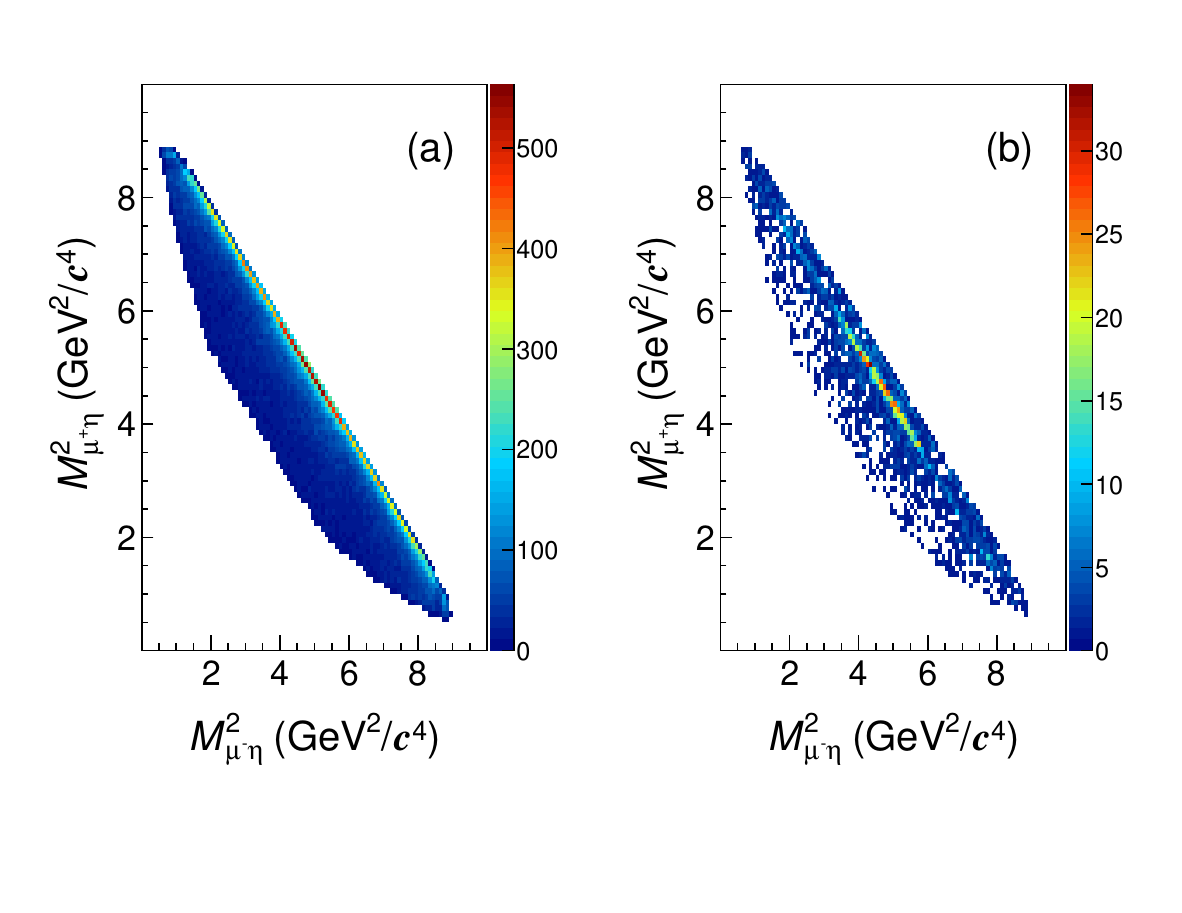}
    \setlength{\abovecaptionskip}{-1pt}
    \caption{Dalitz plots from (a) the signal 
    $J/\psi \to \mu^{+}\mu^{-}\eta$ MC sample and (b) the data.
    The color scale in each panel represents the corresponding event density.
    The diagonal enhancement near the kinematic boundary in the signal MC is mainly associated with the concentration of events at low $M_{\mu^+\mu^-}^2$, while the prominent diagonal band in the data is dominated by the $J/\psi \to \rho(\to \pi^+\pi^-)\eta$ background.
    }
    \label{fig:dalitz}
\end{figure}

The Dalitz plots of $M^2_{\mu^+\eta}$ versus $M^2_{\mu^-\eta}$ after applying all the above selection criteria are shown in Fig.~\ref{fig:dalitz}.

\section{Background analysis}
An inclusive MC sample of $8.9 \times 10^9$ $J/\psi$ events is used to investigate potential background processes.
The topological analysis identifies signal events from $J/\psi \to \omega\eta$, with $\omega \to \mu^+\mu^-$ and $\eta \to \gamma\gamma$, as well as background events arising from pions misidentified as muons. Among the surviving background events, the dominant contribution is from the $J/\psi \to \pi^{+}\pi^{-}\eta$ decay, including contributions via the intermediate states $\rho(770)^0$, $\omega(782)$, and $a_2(1320)^{\pm}$.
A partial wave analysis is performed for the $J/\psi \to \pi^{+}\pi^{-}\eta$ decay, 
and the result is used to generate an exclusive MC sample.
Another background contribution arises from the $J/\psi \to \gamma\eta^{\prime} (\to \pi^{+}\pi^{-}\gamma)$ decay. 
The remaining background contributions are collectively classified as “other background”.    
Potential QED background contributions at the $J/\psi$ resonance are estimated using a data sample collected at $\sqrt{s}=3.773~{\rm GeV}$, corresponding to an integrated luminosity of 20.3 fb$^{-1}$.
No events survive under the same selection criteria, indicating that the QED background contribution is negligible.

\section{BF measurement}
\begin{figure}[htbp]
   \centering
   \includegraphics[width=0.48\textwidth,
    trim={0cm 1.53cm 0cm 0cm},
    clip 
    ] {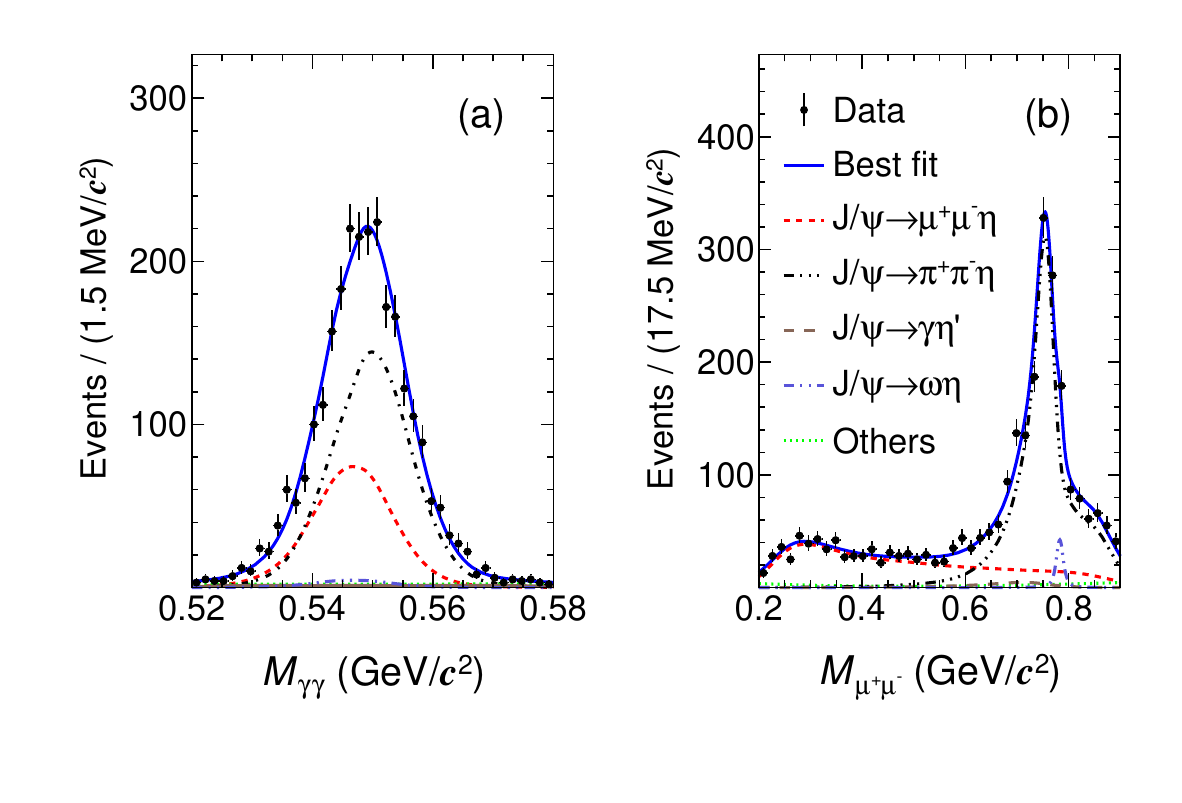}
    \setlength{\abovecaptionskip}{-1.5pt}
    \caption{Two-dimensional fit projections onto (a) $\mathrm{M}_{\gamma\gamma}$ and (b) $\mathrm{M}_{\mu^+ \mu^-}$ distributions. The dots with error bars represent the data, and the blue solid line represents the best fit. The individual fit components are labeled in the legend.}
    \label{fig:eeRec_fit_data}
\end{figure}

The signal yield is determined from a two-dimensional (2D) unbinned maximum-likelihood fit to the joint distribution of $M_{\gamma\gamma}$ and $M_{\mu^+\mu^-}$.
In the fit, the shapes of the signal and the $J/\psi \to \pi^{+}\pi^{-}\eta$ and $J/\psi \to \gamma\eta^{\prime} (\to \pi^{+}\pi^{-}\gamma)$ backgrounds are taken from from the corresponding exclusive MC samples. 
The “other background” component is modeled using a first-order Chebyshev polynomial in $M_{\gamma\gamma}$ and a third-order Chebyshev polynomial in $M_{\mu^+\mu^-}$.
The yield of the $J/\psi \to \omega (\to \mu^+ \mu^-) \eta$ signal component is fixed to $50\pm6$ events, estimated from the BF reported by the PDG~\cite{pdg},
while the yields of the other components are determined by the fit.
Due to the narrow width of the $\omega$ and its small contribution, the interference between this component and the nonresonant $J/\psi \to \mu^+ \mu^- \eta$ decay is ignored in this analysis.
Figure~\ref{fig:eeRec_fit_data} shows the projections of the 2D fit onto the $M_{\gamma \gamma}$ and $M_{\mu^+ \mu^-}$ distributions. 
The signal yield of $J/\psi \to \mu^+ \mu^- \eta$ events is determined to be $N_{\mathrm{sig}} = 843\pm 39$. 
The statistical significance is found to be greater than 10$\sigma$, as determined using a likelihood-ratio test comparing the nominal fit with an alternative fit in which the signal yield is fixed to zero.

The BF of $J/\psi\to\mu^+\mu^-\eta$ is calculated as
\begin{equation}
\label{equ:branch}
   \mathcal{B}(J/\psi \to \mu^+ \mu^- \eta) = \dfrac{ N_{\rm sig} }{ N_{J/\psi} \cdot \varepsilon_{\rm sig}  \cdot \mathcal{B}(\eta\to\gamma\gamma)},
\end{equation}
where $N_{J/\psi}$ = (8.998 $\pm$ 0.039)$\times 10^{9}$ is the number of $J/\psi$ events, 
$\varepsilon_{\rm sig}= 7.8\%$ is the detection efficiency obtained from the signal MC sample,
and $\mathcal{B}(\eta\to\gamma\gamma)=(39.36 \pm 0.18)\%$ is the BF of $\eta \to \gamma \gamma$ taken from the PDG ~\cite{pdg}. 
According to Eq.~(\ref{equ:branch}), the BF of $J/\psi \to \mu^+ \mu^- \eta$ is measured to be $ (3.05 \pm 0.14_{\rm stat.} \pm 0.13_{\rm syst.}) \times 10^{-6}$, 
where the first uncertainty is statistical and the second is systematic.
The systematic uncertainties are discussed in Section~\ref{system}.

Using the world average value of $\mathcal{B}(J/\psi \to \gamma\eta) = (1.090 \pm 0.013) \times 10^{-3}$ reported by the PDG~\cite{pdg}, 
the ratio of the BF of $J/\psi \to \mu^+ \mu^- \eta$ to that of $J/\psi \to \gamma\eta$ is determined to be
\begin{equation}
\label{equ:branch1}
R_{1} \equiv \frac{\mathcal{B}(J/\psi \to \mu^+\mu^-\eta)}{\mathcal{B}(J/\psi\to \gamma\eta)} = (2.80 \pm 0.18) \times 10^{-3}. 
\end{equation}
Using the value $\mathcal{B}(J/\psi \to e^+e^-\eta) = (1.39\pm 0.06_{\rm stat.} \pm 0.07_{\rm syst.}) \times 10^{-5}$ previously measured by BESIII~\cite{BAM:266}, 
the BF ratio between the muonic and electronic EM Dalitz decays is given by
\begin{equation}
\label{equ:branch2}
R_{2} \equiv \frac{\mathcal{B}(J/\psi \to \mu^+\mu^-\eta)}{\mathcal{B}(J/\psi \to e^+e^-\eta)} = 0.219 \pm 0.020,
\end{equation}
where the systematic uncertainties associated with the $\eta$ meson cancel.
$R_2$ is not expected to be unity even if lepton flavor universality holds,
due to the different lepton masses and the corresponding differences in the QED terms and phase space factors. 
 Within the Standard Model framework, the VMD model predicts $R_{2}= 0.248\pm0.012$~\cite{ref:fu}, which is consistent with our measured value within $1.2\sigma$.
The uncertainties quoted in Eqs.~(\ref{equ:branch1}) and (\ref{equ:branch2}) are total uncertainties, including both statistical and systematic contributions.

\section{Systematic Uncertainties}
\label{system}
The systematic uncertainties in the BF measurement originate from the $\mu^{\pm}$ tracking and PID, the photon detection, 
the kinematic fit, 
the signal model, 
the shape of the other background,
the $J/\psi \to \omega \eta$ yield,
the $\mathcal{B}(\eta \to \gamma \gamma)$ taken from the PDG, 
the number of $J/\psi$ events,
and the MC statistics.

The systematic uncertainty from the $\mu^{\pm}$ tracking is assigned to be 1.0\% per track~\cite{sys:helix}.
The systematic uncertainty from the $\mu^{\pm}$ PID is investigated using the control sample $e^+e^-\rightarrow \gamma\mu^+\mu^-$ at $\sqrt{s}=3.097$ GeV.
The $\mu^{\pm}$ PID efficiencies in the control sample are weighted according to the 2D momentum and polar-angle distributions to match those in the signal decay $J/\psi\rightarrow \mu^+\mu^-\eta$.
The difference in the $\mu^{\pm}$ PID efficiencies between data and MC simulation is 1.5\% per track.  

The systematic uncertainty in the photon detection efficiency has been studied using a control sample of $J/\psi\rightarrow\pi^+\pi^-\pi^0$~\cite{sys:Ref2} events,
and is determined to be 0.5\% per photon in the EMC barrel and 1.5\% per photon in the EMC end caps. 
After weighting the uncertainties according to the photon polar-angle distributions in the signal MC sample, the uncertainty associated with the reconstruction of the two photons is estimated to be $1.1\%$.

The systematic uncertainty of the kinematic fit arises from the difference in the track-helix parameters between data and MC simulation.
Following the same method described in Ref.~\cite{sys:Ref1}, 
the resulting difference in the detection efficiencies with and without the helix-parameter correction,
0.4\%, is assigned as the systematic uncertainty.

In the nominal analysis, the signal MC sample is generated using a TFF modeled in the single-pole approximation with the pole mass $\Lambda = (2.56 \pm 0.04_{\rm stat.} \pm 0.03_{\rm syst.})$~GeV/$c^{2}$~\cite{BAM:266}. Alternative signal MC samples are generated by varying $\Lambda$ by $\pm1\sigma$ and by adopting a modified multipole TFF that includes an additional $\rho$-resonance contribution.
The maximum difference in the detection efficiency, 1.2\%, is taken as the systematic uncertainty from the signal model.

The systematic uncertainty from the shape of the other background is evaluated by replacing the Chebyshev polynomial with the shape extracted from the inclusive MC sample.
The resulting difference in the BF measurement, 0.9\%, is assigned as the corresponding systematic uncertainty.

The systematic uncertainty due to the $J/\psi\to\omega\eta$ yield is examined by randomly sampling its BF $10^3$ times from a Gaussian distribution whose width equals its uncertainty and repeating the fit. The distribution of the relative difference in the signal yield is fitted with a Gaussian function,
and its width of 0.1\% is taken as the systematic uncertainty.

The uncertainty due to the quoted BF of $\eta\rightarrow\gamma\gamma$~\cite{pdg} is 0.5\%, 
and the uncertainty in the number of $J/\psi$ events is 0.4\%~\cite{Ablikim:Jpsi}.
The systematic uncertainty due to the MC statistics is calculated by $\sqrt{(1-\varepsilon_{\rm sig}) / (\varepsilon_{\rm sig} \cdot N_{\mathrm{gen}})}$ with $N_{\rm gen} = 9.025\times 10^{5}$, giving 0.4\%.

Table~\ref{eemumusys} summarizes the above systematic uncertainties. 
Assuming all sources to be independent, we obtain a total systematic uncertainty of 4.2\% by adding them in quadrature.

\begin{table}[h]
    \centering
    \caption{\small Relative systematic uncertainties in the BF measurement of $J/\psi \to \mu^+ \mu^- \eta$.}
    \renewcommand{\arraystretch}{1.2}
    \setlength{\extrarowheight}{0.9ex}
    \setlength{\tabcolsep}{3.5pt}
    \renewcommand{\arraystretch}{1.0}
    \vspace{0.20cm}
    \begin{tabular}{ccl}
            \hline\hline
            Source &  Uncertainty (\%) \\
            \hline
            $\mu^{\pm}$ tracking                          &$2.0$   \\
            $\mu^{\pm}$ PID                                & $3.0$ \\
            Photon detection                              &$1.1$  \\ 
            Kinematic fit                              &$0.4$ \\
            Signal model                                & $1.2$  \\
            Other background shape                      & $0.9$    \\
            $J/\psi\to\omega\eta$ yield & 0.1 \\
            $\mathcal{B}(\eta\rightarrow\gamma\gamma)$ & $0.5$ \\
            Number of $J/\psi$ events                 &   $0.4$  \\
            MC statistics                            &   $0.4$\\
            \hline
            Total & $4.2$\\
            \hline\hline
   \end{tabular}%
   \label{eemumusys}
\end{table}

\section{Summary}
 We have observed the EM Dalitz decay $J/\psi \to \mu^+ \mu^- \eta$ for the first time using a sample of (8.998 $\pm$ 0.039) $\times$ 10$^9$ $J/\psi$ events~\cite{Ablikim:Jpsi}, with the $\eta$ meson reconstructed via $\eta \to \gamma \gamma$. The statistical significance is greater than 10$\sigma$. The measured BF is $(3.05\pm 0.14_{\rm stat.} \pm 0.13_{\rm syst.}) \times 10^{-6}$,
which agrees with the theoretical prediction of $(3.0\pm 0.1)\times 10^{-6}$~\cite{ref:fu} based on the VMD model in the single-pole approximation.
Using the BF of the radiative decay $J/\psi \to \gamma\eta$, we determine the ratio $ \mathcal{B}(J/\psi \to \mu^+\mu^-\eta) / \mathcal{B}(J/\psi \to \gamma\eta)$ to be $(2.80 \pm 0.18) \times 10^{-3}$,
consistent with the VMD expectation of $2.74\times 10^{-3}$~\cite{ref:21} within $1\sigma$. 
Furthermore, the ratio of the BFs of the muonic and electronic EM Dalitz decay modes is determined to be $0.219 \pm 0.020$.
This result is consistent with the VMD prediction of $0.248\pm 0.012$~\cite{ref:fu} within $1.2\sigma$, showing no evidence of lepton flavor universality violation.
Due to the limited statistics, the TFF cannot yet be extracted precisely from the $J/\psi\to\mu^{+}\mu^{-}\eta$ decay.
Nevertheless, this analysis provides a basis for further studies of the EM structure of the $J/\psi \to \eta$ transition at the future Super Tau-Charm Facility~\cite{Achasov_2023}.
  
\section{Acknowledgement}
The BESIII Collaboration thanks the staff of BEPCII (https://cstr.cn/31109.02.BEPC) and the IHEP computing center for their strong support. This work is supported in part by National Key R\&D Program of China under Contracts Nos. 2025YFA1613900, 2023YFA1606000, 2023YFA1606704;
Outstanding Youth Foundation of Henan Province No. 252300421219; 
National Natural Science Foundation of China (NSFC) under Contracts Nos. 12305102, 12275067, 12275068, 12105077, 11635010, 11935015, 11935016, 11935018, 12025502, 12035009, 12035013, 12061131003, 12192260, 12192261, 12192262, 12192263, 12192264, 12192265, 12221005, 12225509, 12235017, 12342502, 12361141819, 12535005; the Chinese Academy of Sciences (CAS) Large-Scale Scientific Facility Program; the Strategic Priority Research Program of Chinese Academy of Sciences under Contract No. XDA0480600; CAS under Contract No. YSBR-101; 100 Talents Program of CAS; The Institute of Nuclear and Particle Physics (INPAC) and Shanghai Key Laboratory for Particle Physics and Cosmology;
Science and Technology R$\&$D Program Joint Fund Project of Henan Province No.225200810030, Science and Technology Innovation Leading Talent Support Program of Henan Province No. 254000510039; 
Natural Science Foundation of Henan Province Nos. 252300421491, 242300420250; 
Agencia Nacional de Investigaci\'on y Desarrollo de Chile (ANID), Chile under Contract No. ANID CCTVal CIA250027; Istituto Nazionale di Fisica Nucleare, Italy; Knut and Alice Wallenberg Foundation under Contracts Nos. 2021.0174, 2021.0299, 2023.0315; Ministry of Development of Turkey under Contract No. DPT2006K-120470; National Research Foundation of Korea under Contract No. RS-2026-25486791; National Science and Technology fund of Mongolia; Polish National Science Centre under Contracts Nos. 2024/53/B/ST2/00975, 2025/58/E/ST2/00051; STFC (United Kingdom); Swedish Research Council under Contract No. 2019.04595; U. S. Department of Energy under Contract No. DE-FG02-05ER41374.

\bibliographystyle{apsrev4-2}

\begin{thebibliography}{0}%
\makeatletter
\providecommand \@ifxundefined [1]{%
 \@ifx{#1\undefined}
}%
\providecommand \@ifnum [1]{%
 \ifnum #1\expandafter \@firstoftwo
 \else \expandafter \@secondoftwo
 \fi
}%
\providecommand \@ifx [1]{%
 \ifx #1\expandafter \@firstoftwo
 \else \expandafter \@secondoftwo
 \fi
}%
\providecommand \natexlab [1]{#1}%
\providecommand \enquote  [1]{``#1''}%
\providecommand \bibnamefont  [1]{#1}%
\providecommand \bibfnamefont [1]{#1}%
\providecommand \citenamefont [1]{#1}%
\providecommand \href@noop [0]{\@secondoftwo}%
\providecommand \href [0]{\begingroup \@sanitize@url \@href}%
\providecommand \@href[1]{\@@startlink{#1}\@@href}%
\providecommand \@@href[1]{\endgroup#1\@@endlink}%
\providecommand \@sanitize@url [0]{\catcode `\\12\catcode `\$12\catcode
  `\&12\catcode `\#12\catcode `\^12\catcode `\_12\catcode `\%12\relax}%
\providecommand \@@startlink[1]{}%
\providecommand \@@endlink[0]{}%
\providecommand \url  [0]{\begingroup\@sanitize@url \@url }%
\providecommand \@url [1]{\endgroup\@href {#1}{\urlprefix }}%
\providecommand \urlprefix  [0]{URL }%
\providecommand \Eprint [0]{\href }%
\providecommand \doibase [0]{https://doi.org/}%
\providecommand \selectlanguage [0]{\@gobble}%
\providecommand \bibinfo  [0]{\@secondoftwo}%
\providecommand \bibfield  [0]{\@secondoftwo}%
\providecommand \translation [1]{[#1]}%
\providecommand \BibitemOpen [0]{}%
\providecommand \bibitemStop [0]{}%
\providecommand \bibitemNoStop [0]{.\EOS\space}%
\providecommand \EOS [0]{\spacefactor3000\relax}%
\providecommand \BibitemShut  [1]{\csname bibitem#1\endcsname}%
\let\auto@bib@innerbib\@empty
\end{thebibliography}%


\begin{thebibliography}{**}
\bibitem{Landsberg:1985fd}
L.~G.~Landsberg, 
\href{https://doi.org/10.1016/0370-1573(85)90129-2}{Phys.\ Rept.\  {\bf 128}, 301 (1985).}

\bibitem{Kroll:1955zu}
N.~M.~Kroll and W.~Wada, 
\href{https://doi.org/10.1103/PhysRev.98.1355}{Phys.\ Rev.\ {\bf 98}, 1355 (1955).}

\bibitem{Achasov:1992ku}
N.~N.~Achasov and A.~A.~Kozhevnikov,
\href{https://doi.org/10.1142/S0217751X92002180}{  Sov.\ J.\ Nucl.\ Phys.\  {\bf 55}, 449 (1992).}

\bibitem{Klingl:1996by}
F.~Klingl, N.~Kaiser and W.~Weise,
\href{https://doi.org/10.1007/s002180050167}{Z.\ Phys.\ A {\bf 356}, 193 (1996).}

\bibitem{Faessler:1999de}
A.~Faessler, C.~Fuchs and M.~I.~Krivoruchenko,
\href{https://doi.org/10.1103/PhysRevC.61.035206}{Phys.\ Rev.\ C {\bf 61}, 035206 (2000).}

\bibitem{Terschluesen:2010ik}
C.~Terschlusen and S.~Leupold,
\href{https://doi.org/10.1016/j.physletb.2010.06.033}{Phys.\ Lett.\ B {\bf 691}, 191 (2010).}

\bibitem{Ivashyn:2011hb}
S.~Ivashyn,
\href{https://doi.org/10.48550/arXiv.1111.1291}{Prob.\ Atomic Sci.\ Technol.\  {\bf 2012N1}, 179 (2012).}

\bibitem{GellMann:1961tg}
M.~Gell-Mann and F.~Zachariasen,
\href{https://doi.org/10.1103/PhysRev.124.953}{Phys.\ Rev.\  {\bf 124}, 953 (1961).}

\bibitem{Bauer:1977iq}
T.~H.~Bauer, R.~D.~Spital, D.~R.~Yennie and F.~M.~Pipkin,
\href{https://doi.org/10.1103/RevModPhys.50.261}{Rev.\ Mod.\ Phys.\  {\bf 50}, 261 (1978).}

\bibitem{ref:fu}
J.~L.~Fu, H.~B.~Li, X.~S.~Qin, and M.~Z.~Yang,
\href{https://www.worldscientific.com/doi/abs/10.1142/S0217732312502239}{Mod.\ Phys.\ Lett.\ A {\bf 27}, 1250223 (2012).}

\bibitem{pdg} F. Takahashi  {\it et al.} (Particle Data Group), 
\href{https://pdg.lbl.gov/index.html}{Int.\ J.\ Mod.\ Phys.\ A {\bf 41}, 2630011 (2026)}.

\bibitem{pi0_to_gll_A2} 
S. Prakhov \textit{et al.} (A2 Collaboration at MAMI), \href{https://journals.aps.org/prc/abstract/10.1103/9tk1-2dxp}{Phys.\ Rev.\ C {\bf 113}, 065202 (2026)}. 


\bibitem{eegamma_A2} 
P. Adlarson \textit{et al.} (A2 Collaboration at MAMI), \href{https://journals.aps.org/prc/abstract/10.1103/PhysRevC.95.035208}{Phys.\ Rev.\ C {\bf 95}, 035208 (2017)}.

\bibitem{mumugamma_NA60}
R. Arnaldi et al. (NA60 Collaboration), 
\href{https://doi.org/10.1016/j.physletb.2009.05.029}{Phys.\ Lett.\ B {\bf 677}, 260 (2009)}. 

\bibitem{mumugamma1_NA60}
R. Arnaldi et al. (NA60 Collaboration),
\href{https://doi.org/10.1016/j.physletb.2016.04.013}{Phys.\ Lett.\ B {\bf 757}, 437 (2016)}.

\bibitem{etaptoeegamma}
M.~Ablikim {\it et al.} (BESIII Collaboration), \href{https://doi.org/10.1016/j.physletb.2009.05.029}{Phys.\ Rev.\ D {\bf 92}, 012001 (2009)}; 
M.~Ablikim {\it et al.} (BESIII Collaboration), \href{https://journals.aps.org/prd/abstract/10.1103/PhysRevD.109.072001}{Phys.\ Rev.\ D {\bf 109}, 072001 (2024)}.

\bibitem{pi0}
A. Anastasi {\it et al}. (KLOE-2 Collaboration), 
\href{https://doi.org/10.1016/j.physletb.2016.04.015}{Phys.\ Lett.\ B {\bf 757}, 362 (2016)}.

\bibitem{eta1}
D. Babusci {\it et al}. (KLOE-2 Collaboration), 
\href{https://doi.org/10.1016/j.physletb.2015.01.011}{Phys.\ Lett.\ B {\bf 742}, 1 (2015)}.



\bibitem{BAM:266}
M.~Ablikim {\it et al.} (BESIII Collaboration),
  \href{https://doi.org/10.1103/PhysRevD.104.099901}{Phys.\ Rev.\ D {\bf 104}, 099901 (2021).}

\bibitem{eta1405}
M.~Ablikim {\it et al.} (BESIII Collaboration),
\href{https://doi.org/10.1103/PhysRevD.109.012007}{Phys.\ Rev.\ D {\bf 109}, 012007 (2024).}

\bibitem{X1835}
M.~Ablikim {\it et al.} (BESIII Collaboration),
\href{https://journals.aps.org/prl/abstract/10.1103/PhysRevLett.129.022002}{Phys.\ Rev.\ Lett {\bf 129}, 022002 (2022).}

\bibitem{hc}
M.~Ablikim {\it et al.} (BESIII Collaboration),
\href{https://doi.org/10.1103/PhysRevD.110.L111101}{Phys.\ Rev.\ D {\bf 110}, L111101 (2024).}

\bibitem{etac}
M.~Ablikim {\it et al.} (BESIII Collaboration), \href{https://journals.aps.org/prd/abstract/10.1103/PhysRevD.106.112002}{Phys.\ Rev.\ D {\bf 106}, 112002 (2022)}.

 \bibitem{xcj}
M.~Ablikim {\it et al.} (BESIII Collaboration),
\href{https://journals.aps.org/prl/abstract/10.1103/PhysRevLett.118.221802}{Phys.\ Rev.\ Lett {\bf 118}, 221802 (2017)}.

\bibitem{D0}
M.~Ablikim {\it et al.} (BESIII Collaboration), \href{https://journals.aps.org/prd/abstract/10.1103/PhysRevD.104.112012}{Phys.\ Rev.\ D {\bf 104}, 112012 (2021)}.
  
\bibitem{x3872_to_jpsimm_LHCb}
R. Aaij {\it et al}. (LHCb Collaboration), 
\href{https://journals.aps.org/prd/pdf/10.1103/jqvq-sbp7}{Phys.\ Rev.\ D {\bf 113}, 072019 (2026)}.

\bibitem{Arnaldi:2009aa}
R.~Arnaldi \textit{et al.} (NA60 Collaboration),
\href{https://doi.org/10.1016/j.physletb.2009.05.029}{Phys. Lett. B \textbf{677}, 260 (2009).}

\bibitem{Arnaldi:2009aa1}
R.~Arnaldi \textit{et al.} (NA60 Collaboration),
\href{https://doi.org/10.1016/j.physletb.2016.04.013}{Phys. Lett. B \textbf{757}, 437 (2016).}

\bibitem{tershlusen_2012} 
C. Terschl{\rm ${\rm \ddot{u}}$}sen, S. Leupold, and M. Lutz, 
\href{https://link.springer.com/article/10.1140/epja/i2012-12190-6}{Eur. Phys. J. A {\bf 48}, 190 (2012).}

\bibitem{schneider_2012} 
S. P. Schneider, B. Kubis, and F. Niecknig, 
\href{https://journals.aps.org/prd/abstract/10.1103/PhysRevD.86.054013}{Phys. Rev. D {\bf 86}, 054013 (2012).}

\bibitem{BAM:325}
M.~Ablikim {\it et al.} (BESIII Collaboration),
\href{https://journals.aps.org/prd/abstract/10.1103/yglh-sfbg}{Phys.\ Rev.\ D {\bf 112}, L011101 (2025).}

\bibitem{ref:21}
J.~L.~Zhang, J.~J.~He, T.~J.~Zhu, S.~Q.~Xu, and R.~Wang,
\href{https://doi.org/10.1142/S0217751X24430188}{Int.\ J.\ Mod.\ Phys.\ A {\bf 34}, 1950129 (2019).}

\bibitem{ref:22}
L.~M.~Gu, H.~B.~Li, X.~X.~Ma, and M.~Z.~Yang,
\href{https://doi.org/10.1103/PhysRevD.100.016018}{Phys.\ Rev.\ D {\bf 100}, 016018 (2019).}


\bibitem{ref:23}
Y.~H.~Chen, Z.~H.~Guo, and B.~S.~Zou,
\href{https://journals.aps.org/prd/abstract/10.1103/PhysRevD.91.014010}{Phys.\ Rev.\ D {\bf 91}, 014010 (2015).}

\bibitem{ref:24}
L.~W.~Yan, Y.~H.~Chen, C.~G.~Duan, and Z.~H.~Guo,
\href{https://journals.aps.org/prd/abstract/10.1103/PhysRevD.107.034022}{Phys.\ Rev.\ D {\bf 107}, 034022 (2023).}

\bibitem{ref:25}
B.~Kubis, and F.~Niecknig,
\href{https://journals.aps.org/prd/abstract/10.1103/PhysRevD.91.036004}{Phys.\ Rev.\ D {\bf 91}, 036004 (2015).}

\bibitem{Ablikim:Jpsi}
M.~Ablikim \textit{et al.} (BESIII Collaboration),
\href{https://iopscience.iop.org/article/10.1088/1674-1137/ac5c2e}{Chin. Phys. C \textbf{46}, 074001 (2022).}


\bibitem{24}
M.~Ablikim {\it et al.} (BESIII Collaboration),
\href{https://doi.org/10.1016/j.nima.2009.12.050}{Nucl.\ Instrum.\ Meth.\ A {\bf 614}, 345 (2010).}

\bibitem{25}
C.~H.~Yu {\it et al.},
\href{https://doi.org/10.18429/JACoW-IPAC2016-TUYA01}{in 7th International Particle Accelerator Conference (2016) p. TUYA01.}

\bibitem{26}
M.~Ablikim {\it et al.} (BESIII Collaboration),
\href{https://doi.org/10.1088/1674-1137/44/4/040001}{Chin. Phys. C {\bf 44}, 040001 (2020).}

\bibitem{27}
J.~Lu, Y.~Xiao, X.~Ji,
\href{https://doi.org/10.1007/JHEP01(2025)089}{Radiat. Detect. Technol. Methods {\bf 4}, 337 (2020).}

\bibitem{28}
J.~W.~Zhang, L.~H.~Wu, S.~S.~Sun {\it et al.},
\href{https://arxiv.org/pdf/2410.20063v2}{ Radiat. Detect. Technol. Methods {\bf 6}, 289 (2022).}
 
\bibitem{etof1} 
X.~Li {\it et al.}, 
\href{https://link.springer.com/article/10.1007/s41605-017-0014-2}{Radiat. Detect. Technol. Methods {\bf{1}}, 13 (2017)};
Y.~X.~Guo {\it et al.}, 
\href{https://link.springer.com/article/10.1007/s41605-017-0012-4}{Radiat. Detect. Technol. Methods {\bf{1}}, 15 (2017)};
P.~Cao {\it et al.}, 
\href{https://www.sciencedirect.com/science/article/pii/S0168900219314068}{Nucl.~Instrum.~Meth.~{\bf A} {\bf{953}}, 163053 (2020).} 

\bibitem{geant4}
S.~Agostinelli {\it et al.} (GEANT4 Collaboration), \href{https://doi.org/10.1016/S0168-9002(03)01368-8}{ Nucl.\ Instrum.\ Meth.\ A {\bf 506}, 250 (2003).}

\bibitem{ref:kkmc}
S.~Jadach, B.~F.~L.~Ward and Z.~Was,
\href{https://doi.org/10.1016/S0010-4655(00)00048-5}{  Phys.\ Rev.\ D {\bf 63}, 113009 (2001);
Comput.\ Phys.\ Commun.\  {\bf 130}, 260 (2000).}

\bibitem{ref:evtgen}
D.~J.~Lange,
\href{https://doi.org/10.1016/S0168-9002(01)00089-4}{  Nucl.\ Instrum.\ Meth.\ A {\bf 462}, 152 (2001);}
R.~G.~Ping,
\href{https://doi.org/10.1088/1674-1137/32/8/001}{ Chin. Phys. C {\bf 32}, 599 (2008).}

\bibitem{ref:lundcharm}
J.~C.~Chen, G.~S.~Huang, X.~R.~Qi, D.~H.~Zhang and Y.~S.~Zhu,
\href{https://doi.org/10.1103/PhysRevD.62.034003}{Phys.\ Rev.\ D {\bf 62}, 034003 (2000);}
R.~L.~Yang, R.~G.~Ping and H.~Chen,
\href{https://doi.org/10.1088/0256-307X/31/6/061301}{Chin.\ Phys.\ Lett.\  {\bf 31}, 061301 (2014).}

\bibitem{photos2}
E.~Barberio, B.~van Eijk and Z.~Was,
\href{https://doi.org/10.1016/j.nuclphysbps.2009.03.009}{Comput. Phys. Commun. {\bf 66}, 115 (1991).}

\bibitem{sys:helix}
M.~Ablikim \textit{et al.} (BESIII Collaboration),
\href{https://doi.org/10.1103/PhysRevD.111.052002}{Phys. Rev. D {\bf 111}, 052002 (2025).}

\bibitem{sys:Ref2}
M.~Ablikim \textit{et al.} (BESIII Collaboration),
\href{https://doi.org/10.1103/PhysRevD.93.039906}{Phys. Rev. D {\bf 93}, 039906 (2016).}

\bibitem{sys:Ref1}
M.~Ablikim \textit{et al.} (BESIII Collaboration),
\href{https://doi.org/10.1103/PhysRevD.87.012002}{Phys. Rev. D {\bf 87}, 012002 (2013).}


\bibitem{Achasov_2023}
M. Achasov \textit{et al.}, 
\href{https://doi.org/10.1007/s11467-023-1333-z}{Front. Phys. {\bf 19}, 14701 (2023).}



\end{thebibliography}

\end{document}